\newcommand{\be}{\begin{equation}}\newcommand{\ee}{\end{equation}}
\newcommand{\bea}{\begin{eqnarray}}\newcommand{\eea}{\end{eqnarray}}

\documentstyle[12pt]{article}

\begin{document}

\thispagestyle{empty}
\hfill JINR E2-92-301 \\
\bigskip\bigskip\begin{center} {\bf
\Large{Boussinesq-type equations from  nonlinear realizations of
$W_3$}}
\end{center}  \vskip 1.0truecm
\centerline{{\bf E. Ivanov, S. Krivonos and R.P. Malik}}
\begin{center}
{\it Laboratory of Theoretical Physics,} \\
{\it Joint Institute for Nuclear Research, Dubna, Head Post Office}\\
{\it P.O. Box 79, 101 000 Moscow, Russia}
\end{center}
\bigskip \nopagebreak \begin{abstract}

We construct new coset realizations of infinite-dimensional linear
$W_3^{\infty}$ symmetry associated with Zamolodchikov's $W_3$ algebra
 which are different from
the previously explored $sl_3$ Toda realization of $W_3^{\infty}$. We
deduce the Boussinesq and modified Boussinesq equations as constraints
on the geometry of the corresponding coset manifolds.
The main characteristic features of these realizations are:
i. Among the coset parameters there are the space and time
coordinates $x$  and $t$  which enter the Boussinesq
equations, all other coset
parameters are regarded as fields depending on these coordinates;
ii. The spin 2 and 3 currents of $W_3$ and two spin 1
$U(1)$ Kac-Moody currents as well as two spin 0 fields related to the $W_3$
currents via Miura maps, come out as the only essential parameters-fields of
these cosets. The remaining coset fields are covariantly expressed through
them;
iii. The Miura maps get a new geometric interpretation as
$W_3^{\infty}$ covariant constraints
which relate the above fields while passing from one
coset manifold to another;
iv. The Boussinesq equation and two kinds of the modified Boussinesq
equations appear geometrically as the dynamical constraints
accomplishing $W_3^{\infty}$ covariant reductions of original coset manifolds
to
their two-dimensional geodesic submanifolds;
v. The zero-curvature representations for these equations arise automatically
as a consequence of the covariant reduction;
vi. $W_3$ symmetry of the Boussinesq equations amounts to the left action of
$W_3^{\infty}$ symmetry on its cosets. The approach proposed could
provide a universal geometric description of the relationship between
$W$-type algebras and integrable hierarchies.

\end{abstract} \bigskip
\begin{center}
Submitted to {\it Int. J. Mod. Phys. A}
\end{center}
\vfill
\begin{center}
Dubna 1992
\end{center}
\newpage\setcounter{page}1

\section{Introduction}

The existence of intimate relationships between $W$-algebras on
one hand and conformal field theories and integrable systems in 1+1
dimensions on the other (see, e.g. [1-12]) is a fairly well-established fact
which has profound and far-reaching implications in modern mathematical
physics, especially in what concerns the string theory and $2D$ gravity.
One of the most exciting discoveries in this area is the understanding of
the property that various $W$ algebras and their superextensions provide the
second Hamiltonian structures for the generalized KdV and KP hierarchies
as well as superextensions of the latter. For instance, the $W_2$ (Virasoro)
algebra defines the second Hamiltonian structure
for the KdV hierarchy [5], $W_3$ for the Boussinesq one [6], $W_{1+\infty}$
for the KP hierarchy [7], etc. A natural description of the correspondence
between the $W$ algebras and $1+1$ integrable hierarchies is achieved
in terms of pseudo-differential operators \cite{a8}. It is
of interest to understand these
remarkable relationships proceeding directly from the
intrinsic geometries of $W$ symmetries. Besides providing new insights
into the geometric origin of the above hierarchies,
this could shed more light on the geometry of the associated theories, such as
$W$ gravities, $W$ strings, etc.

To understand the geometry behind a symmetry group G,
the key concept is to consider it as a group of
transformations acting on the coset space $G/H$ with an appropriately chosen
stability subgroup $H$. This is the content of the famous nonlinear
(or coset) realization method \cite{a13}.
In the papers of two of us (E.I. \& S.K.) \cite {{a12},{a14},{a15}},
it has been argued
that the most direct and fruitful way of revealing geometric features
of $W$ symmetries is via this method.
However, it had been originally invented to treat the Lie type symmetries,
therefore, its application to $W_N$ symmetries encounters
difficulties because
$W_N$ algebras for $N \geq 3$ are not Lie ones. A way out of this difficulty
has been proposed in
ref. \cite{a12}. It consists in replacing $W_N$ algebras
by some associate infinite-dimensional {\it linear} algebras $W_{N}^{\infty}$
which appear if one treats as new independent generators all the
higher-spin composite generators present in the enveloping algebra of the
basic $W_N$ generators (these are the spin 2 and 3 generators in the
$W_3$ case, the spin 2,3 and 4 ones in the $W_4$ case, etc). $W_N$ symmetries
can then be viewed as particular realizations of linear $W_N^{\infty}$
symmetries. To the latter, one may apply the entire arsenal of the
coset realization techniques. The authors of ref.\cite{a12} have
constructed a coset realization of the product of two
light-cone copies of $W_3^{\infty}$ and have shown that after imposing an
infinite number of the inverse Higgs \cite{a16} type covariant constraints on
the relevant Cartan forms (this is called the covariant reduction
of a given coset \cite{{a12},{a14}}), one is
left with the realization of $W_3^{\infty}$
on two scalar 2D fields which coincides with the well-known $sl_3$ Toda
realization of
$W_3$ \cite{{a4},{a10}}. As a consequence of the covariant reduction
constraints, the
scalar fields turned out to satisfy the $sl_3$ Toda lattice field
equations (or their free version, depending on the choice of the vacuum
stability subgroup) which thus had proven to be intimately related to the
intrinsic geometry of $W_3^{\infty}$. An analogous treatment of the $sl_2$
Toda system (Liouville theory) in the framework of much simpler
nonlinear realization of $W_2$ (Virasoro) symmetry has been earlier given in
\cite{a14}
\footnote{$W_2$ symmetry is linear, so there is no direct necessity to
pass to something like $W_2^{\infty}$ while constructing its nonlinear
realization. However, this necessity comes out if one tries to understand
the KdV hierarchy in the nonlinear realization approach \cite{a17}.}.

In the present paper we construct coset realizations of
$W_3^{\infty}$ (its one copy) which are {\it different} from those found in
\cite{a12}. We demonstrate that there exists a set of the covariant
reduction constraints which reduces the
number of independent coset parameters-fields
to the two fields of conformal spins 2 and 3 identified with the
currents of $W_3$ and simultaneously gives rise
to the Boussinesq equation for these fields. Both
the spatial $(x)$ and
evolution $(t)$ coordinates of these fields naturally appear as
the parameters of the coset considered. The generator to which $t$ is
attached coincides with the Hamiltonian used in the standard Hamiltonian
approach to the Boussinesq equation, thus establishing a link between
the second Hamiltonian structure for this equation \cite{a6} and our
geometric approach. The
Miura map for the Boussinesq equation gets also a simple geometric
interpretation. One
enlarges the coset by adding two spin 1
generators from the stability subgroup and then imposes
additional covariant constraints which covariantly express the spin 2 and
spin 3 coset fields in terms of the two new spin 1 fields. The resulting
expressions are just the Miura transformations relating the Boussinesq
equation to a ``modified'' Boussinesq equation. Thus the Miura map arises as
a manifestly covariant
relation between parameters of a coset of $W_3^{\infty}$. Quite analogously
one may construct further Miura map onto the two spin 0 fields which leads
to the Feigin-Fuchs type representation for the spin 2 and spin 3 fields. One
should transfer two spin 0 generators from the stability group to the coset
(the remaining generators still form a subalgebra) and impose appropriate
inverse Higgs constraints. The zero-curvature
representation for the ordinary and modified Boussinesq equations,
as well as the relevant matrix
Lax pairs, appear very naturally in this picture, basically as the
Maurer-Cartan equations for the reduced cosets. The $W_3$ symmetry of the
Boussinesq equations established recently in \cite{a18} also
immediately follows, it is recognized as left $W_3^{\infty}$ shifts on the
relevant coset manifolds. The higher-order equations from
the Boussinesq hierarchy can be produced by considering more general cosets
of $W_3^{\infty}$ with
additional evolution parameters associated with the higher-spin generators.

The paper is organized as follows.

In Sect.2 we consider a toy example of the nonlinear realizations of Virasoro
($W_2$) symmetry and show that some essential features of the entire
construction can be seen already in this simplified situation. In particular,
the holomorphic component of the conformal stress-tensor comes out
as a parameter of some coset manifold of $W_2$, the Miura maps amount to
covariant constraints on the coset parameters, etc.

In Sect.3 we recapitulate the basic facts about the linear algebra
$W_3^{\infty}$ following ref. \cite{a12} and list its some subalgebras
which
are utilized while constructing coset realizations of $W_3^{\infty}$ symmetry
in the next Section.

In Sect.4 we construct three coset realizations of $W_3^{\infty}$ and
give necessary technical details (parametrizations of the coset elements,
Cartan forms).

Sect.5 is the central, there we apply the covariant reduction procedure
to the cosets constructed in Sect.4 and show that it results in expressing
an infinite tower of the coset parameters-fields in terms of a few essential
ones: either the spin 2 and spin 3 fields $u$, $v$, or two spin 1 fields
$u_1$, $v_1$, or two spin 0 fields $u_0$, $v_0$. Simultaneously one obtains
dynamical equations for these fields, namely the Boussinesq equation
and two types of the modified Boussinesq equations. We explain the geometric
meaning of the appropriate Miura maps and zero-curvature representations
and make contact with the Hamiltonian formulation.

In Sect.6 we study transformation properties of the Boussinesq equation
under left shifts of $W_3^{\infty}$ on the original coset elements. We
show that in the realization on the  spin 2 and spin 3 fields,
the $W_3^{\infty}$ transformations constitute the $W_3$ symmetry
of the Boussinesq equation revealed in a recent paper \cite{a18}.

\setcounter{equation}{0}

\section{A sketch of nonlinear realizations of Virasoro symmetry}

For reader's convenience and to make more clear what kind of
nonlinear realizations of $W_3$ symmetry we are going to construct, we
first dwell on a simpler case of Virasoro ($W_2$) symmetry. We will
demonstrate here that the holomorphic component of the conformal
stress-tensor can be treated as the coset space parameter corresponding
to a kind of the coset realization of one copy of this symmetry.
The Miura map and the Feigin-Fuchs representation for this component
naturally appear in the framework
of some extended coset spaces of $W_2$ as $W_2$- covariant constraints on
the coset parameters.

We restrict our study, like in ref.\cite{a14}, to the truncated $W_2$
formed by the generators
\be  \label{0.1}
W_2 = \left\{ L_{-1}\;,\;L_0\;,\;L_1\;,
\ldots , L_n\;, \ldots\; \right\} \;\;n\geq -1\;;\;\;\; [L_n,\;L_m]
= (n-m) L_{n+m}\;.
\ee
In what follows we will denote by $W_2$ just this algebra and, depending
on the context, use the same term for the corresponding group of
transformations.

As was observed in ref. \cite{a14}, the standard realization
of $W_2$ as conformal transformations of  $R^{1}$ (or $S^{1}$)
coordinate $x$ can be easily re-derived within the framework of
coset realization method.
It is induced by a left action
of the group associated with the algebra (\ref{0.1}) on the one-dimensional
coset over the subgroup generated by
\be  \label{0.2}
L_0\;,\;L_1\;, \ldots , L_n\;, \ldots; \;\;n\geq 0\;.
\ee
Namely, parametrizing an element of this coset as
$$ g(x) = {\mbox e}^{x L_{-1}}$$
and defining the group action on it following the standard
rules of ref. \cite{a13}
\be \label{0.3}
g_0 \;g(x) = g(x')\; h (x, g_0)\;,\;\;\; g_0 = {\mbox exp}\left(
\sum_{n\geq -1} \lambda_n L_n \right)\;,
\ee
where $h$ is some induced transformation from the stability subgroup, one
obtains for $x$ the standard conformal transformations (regular at the
origin)
\be \label{0.03}
\delta x \equiv \lambda (x) = \sum_{n \geq -1} \lambda_n \;(x)^{n+1}\;.
\ee

The above coset is not reductive in the sense that the coset generator
$L_{-1}$ is rotated by the stability group into the generators of the latter
and this causes some difficulties in applying the standard techniques
of ref. \cite{a13} to the present case.
The simplest reductive coset manifold of $W_2$ is obtained by treating {\it
all}
generators (\ref{0.1}) as the coset ones:
\be \label{0.4}
g(x) \Rightarrow \tilde{g}(x) = {\mbox e}^{x L_{-1}}\cdot
\left( \prod_{n\geq 3}
{\mbox e}^{u_n (x) L_n}\right) \cdot {\mbox e}^{u_1 (x) L_1}\;
{\mbox e}^{u(x) L_2}\; {\mbox e}^{u_0 (x) L_0}\;,
\ee
where the coset parameters are regarded as fields given on the line manifold
$x$,
i.e. as a kind of goldstone fields, and the special arrangement
of factors has been chosen for further convenience. Under the left $W_2$
shifts the coordinate $x$ as before transforms according to the rule
(\ref{0.03}), while
the coset parameters-fields transform through themselves and the function
$\lambda (x)$.
For instance,
\bea
\delta u_0  &=& -\lambda   u'_0  +  \lambda' \nonumber \\
\delta u_1  &=& - \lambda  u'_1   - \lambda' \; u_1 +
\frac{1}{2} \lambda'' \nonumber \\
\delta u  &=& -\lambda  u' -2 \lambda'  u +
\frac{1}{6}  \lambda''' \;,\mbox{ etc.} \label{0.5}
\eea
We observe that $u_0(x)$ transforms as a $2D$ dilaton, while $u(x)$ as
a holomorphic component of conformal stress-tensor. To see that the latter
property is not accidental, let us look at the structure of the Cartan forms
for the nonlinear realization in question.

The Cartan forms are introduced as usual by
\be \label{0.6}
\tilde{g}^{-1} d \tilde{g} = \sum_{n\geq -1} \omega_n L_n
\ee
and are invariant by construction under the left action of $W_2$ symmetry.
They can be easily evaluated using the commutation relations (\ref{0.1}). A
few first ones are as follows
\bea
\omega_{-1} &=& {\mbox e}^{-u_{0}} dx \nonumber \\
\omega_{0} &=& (u'_0 - 2 u_1)\;dx \nonumber \\
\omega_{1} &=& {\mbox e}^{u_{0}}\;( u'_1 + (u_1)^2 - 3 u_2 ) \;dx\;.
\label{0.7}
\eea
Note that the higher-order forms, like $\omega_{0}$ and $\omega_{1}$,
contain the pieces linear in the relevant parameters-fields (beginning with
$u_{3}$). Now, keeping in mind invariance of these forms, one may impose
the manifestly covariant inverse Higgs type \cite{a16} constraints
\be \label{0.8}
\omega_{n} = 0\;,\;\;\; \forall_{n\geq 0}\;,
\ee
which can be looked upon as algebraic equations for expressing the
parameters-fields $u_1\;,\;u\;,\;u_n\;(n\geq 3)$ in terms of
$u_0$ and its derivatives in a way compatible with the transformation
properties (\ref{0.5}). Thus $u_0(x)$ is the only essential coset
parameter-field in the present case. Using eqs.(\ref{0.7}) one finds
the coset fields $u_1$ and $u$ to be expressed by
\be \label{0.9}
u_1 = \frac{1}{2} u'_0\;,\;\;\; u = \frac{1}{6} \left[ \frac{1}{2}
(u'_0)^2 + u''_0 \right] \;.
\ee
We see that $u$ indeed has the standard form of the conformal stress-tensor
for the single scalar field in the Feigin-Fuchs representation (an arbitrary
parameter that is usually present in front of the Feigin-Fuchs term can be
attributed
to a rescaling of $u_0$). Thus we have succeeded in deducing the Feigin-Fuchs
representation for the stress-tensor as a covariant relation between the
parameters of certain coset manifold of $W_2$ symmetry.

The above coset realization of $W_2$ is not unique, there exist other ones,
with less trivial stability subgroups.

The first possibility is to
factorize over a one-dimensional subgroup with the generator $L_0$
\be \label{0.10}
{\cal H}_1 =\left\{ L_0 \right\}\;.
\ee
The
relevant coset element and Cartan forms are obtained simply by putting
$u_0 = 0$ in eqs. (\ref{0.4}), (\ref{0.6}) and (\ref{0.7}). The set of
Cartan forms now consists of those living in the coset
($\omega_{-1}\;, \; \omega_{n}\;\;n\geq 1$) and in the stability subalgebra
($\omega_0 = -2 u_1 dx$). The coset forms now
undergo homogeneous $L_0$ rotations while $\omega_0$ transforms
inhomogeneously.
The only essential coset field in this realization is $u_1$; all others are
expressed in terms of it via the covariant constraints
\be \label{0.11}
\omega_n = 0\;,\;\;\;\; \forall_{n\geq 1}
\ee
For $u$ one obtains now the representation
\be \label{0.12}
u = \frac{1}{3} \left[ (u_1)^2 +  u'_1 \right]\;,
\ee
which is known as the ``Miura map''
for the stress-tensor ($u_1$ is interpreted as a $U(1)$
Kac-Moody current). Thus the Miura map also naturally appears in the
nonlinear realization approach to $W_2$ as a covariant relation between
the parameters of the coset of $W_2$ over the subgroup with the algebra
(\ref{0.10}).

Finally, one may extend the stability group algebra by including
the generator $L_1$
\be \label{0.13}
{\cal H} = \left\{ L_0\;,\;L_1 \right\}
\ee
(further extension is impossible since adding, e.g., the generator $L_2$
would immediately entail adding
an infinite set of  $W_2$ generators and we would return to the
non-reductive case discussed in the beginning of this Sect.). The set of
the associate coset fields starts with $u$ which is independent in
this realization. All higher-order fields are expressed through $u$
by the constraints
\be \label{0.14}
\omega_n = 0\;\;\; \;\;\forall_{n\geq 2}\;,
\ee
which are still closed under the left action of $W_2$.

A few words are of need regarding the geometric meaning of the procedure
of
eliminating higher-order coset fields in the above examples. In all cases,
after imposing the inverse Higgs constraints, we are left with the Cartan
forms on the stability subalgebra and the form $\omega_{-1}$. The associate
generators form subalgebras which are particular cases of
what was called ``the covariant reduction subalgebra'' in ref.
\cite{{a12},{a14},{a15}}.
These are the one-generator subalgebra $\left\{ L_{-1} \right\}$ in the
first example, the two-generator subalgebra $\left\{ L_{-1},\;L_0 \right\}$
in the second example and the algebra $sl(3, R) = \left\{
L_{-1}\;,\;L_0\;,\;L_1
\right\}$ in the third example. The coordinate $x$ parametrizes the
one-dimensional cosets of the corresponding subgroups (``covariant reduction
subgroups'' in the terminology of ref. \cite{{a12},{a14},{a15}}) over the
stability
subgroups while the surviving coset
fields ($u_0(x)$, $u_1(x)$ and $u(x)$, respectively) together with their
derivatives of any order specify embedding of these curves into the original
infinite-dimensional cosets of $W_2$. As was argued in \cite{a19} on a simple
finite-dimensional example, such curves (and two-dimensional hypersurfaces in
the cases considered in ref. \cite{{a12},{a14}} and  Section 5 of the present
paper) form fully geodesic submanifolds in the original coset manifolds.
Thus the role of eqs. (\ref{0.8}), (\ref{0.11}), (\ref{0.14}) is to
single out one-dimensional geodesic submanifolds in the cosets of
$W_2$.

The main points one learns from the above discussion are as follows:

i. The conformal stress-tensor $u(x)$ as well as the $2D$ dilaton field
$u_0(x)$ and the $U(1)$ Kac-Moody current $u_1(x)$ can be given a nice
geometric interpretation as the parameters of coset manifolds of
the truncated Virasoro ($W_2$) symmetry (\ref{0.1}).

ii. The free-field Feigin-Fuchs type representation and Miura map
for the stress-tensor appear in a geometric way as covariant constraints
on the parameters of these cosets. These serve to single out
geodesic curves in the original manifolds.

In the examples considered here the inverse Higgs constraints are purely
kinematic, they do not imply any dynamics for the involved fields. To gain a
dynamics, one should arrange the fields to depend, besides $x$, on an
evolution parameter, a time coordinate. One way to do this is to add one
more copy of $W_2$ and to interpret the coset parameters associated with the
two commuting generators $L_{-1}$ as the light-cone $2D$ Minkowski
coordinates.
A nonlinear realization of such a symmetry generalizing first of the
examples considered here has been studied in ref. \cite{a14}.
It has been shown that the relevant inverse Higgs constraints not only
serve to eliminate higher-order coset fields in terms of the $2D$ dilaton,
but also give rise to the dynamical equations for the latter, in particular
to the Liouville equation. Such a version of the inverse Higgs procedure
has been called ``the covariant reduction''. An analogous construction for
two light-cone copies of $W_3$ symmetry has been given in ref.\cite{a12} and
it has led to a
new geometric interpretation of the $sl_3$ Toda theory within the covariant
reduction approach. In the next Sections we shall demonstrate that there
exists another way to incorporate the evolution parameter into a nonlinear
realization of $W_3$ symmetry without doubling the algebra. The relevant
cosets of $W_3$ are a more or less direct generalization of those considered
here. However, due to the presence of the time coordinate, it will turn out
to be possible
to obtain dynamical equations for the coset fields from the
covariant reduction procedure: the Boussinesq and modified
Boussinesq equations. But before
discussing this we need to recall how to apply the notions of
nonlinear realizations to the algebra $W_3$. As was shown in ref.\cite{a12},
the only conceivable way to do this is to deal, instead of $W_3$, with some
infinite-dimensional linear algebra $W_3^{\infty}$ closely related to $W_3$.

\setcounter{equation}{0}

\section{$W_3^{\infty}$ and its subalgebras}

In this Section we shall briefly recall salient features of the algebra
$W_3^{\infty}$ and its relation to $W_3$, closely following ref.\cite{a12}.
To avoid a possible confusion, we point out that, like in \cite{a12},
we start with the most general classical $W_3$ algebra possessing an
arbitrary central charge. Its commutation relations can be found, e.g., in
ref.\cite{a6} (see eqs. (3.1) and (3.2) below).

The central idea invoked in \cite{a12} is to construct a linear algebra
$W_3^{\infty}$
from the nonlinear $W_{3}$ by treating as independent all the higher-spin
composite generators which appear while considering successive commutators
of the basic (spin 2 and 3) $W_3$ generators.

Let us consider the defining relations of the
classical $W_3$ algebra \cite{a6}
\footnote{For correspondence with our
forthcoming work on $N=2$ super $W_3$ algebra \cite{a190} we use a bit
different
normalizations of the spin 3 and spin 4 generators $J_n$
and $J_n^{(4)}$ compared to those used in \cite{a12}. These are the same as in
ref. \cite{a20}}
\bea
\left[ L_n, L_m \right] & = & (n-m)L_{n+m} +
\frac{c}{12}(n^3-n)\delta_{n+m,0} \nonumber \\
\left[ L_n, J_m \right] & = & (2n-m)J_{n+m} \nonumber \\
\left[ J_n, J_m \right] & = & 16 (n-m)J_{n+m}^{(4)}-
   \frac{8}{3} (n-m) \left( n^2 +m^2 -\frac{1}{2}nm-
4 \right)L_{n+m} - \nonumber \\
        & & -\frac{c}{9}(n^2-4)(n^2-1)n\delta_{n+m,0} \quad , \label{1.1}
\eea
where
\be \label{1.2}
J_n^{(4)} = - \frac{8}{c}\sum_m L_{n-m}L_m.
\ee
The structure relations of $W_3^{\infty}$ are then defined as the full set
of commutators between $L_n$, $J_n$, $J_n^{(4)}$ and all higher-spin
composites $J^{(s)}_{n}\; (s\geq 5)$ which come out in the
commutators of
lower-spin generators. All the composite generators, beginning with
$J_n^{(4)}$, are treated as new independent ones.
Thus $W_3^{\infty}$ is formed by an infinite set of generators
$$ L_n,\;J_n,\;J_n^{(4)},\;J_n^{(s)},\;\;s\geq 5 .$$
It should be pointed out that the full set of commutation relations of
$W_3^{\infty}$, as is clear from the above definition,  can be entirely
deduced from the basic $W_3$ relations (3.1), (3.2). For our purposes
here it will be of no need to know the detailed structure of these
commutation relations.

It will be of importance that the central
charge $c$ is non-zero in (3.1). The presence of this parameter strongly
influences the structure of $W_3^{\infty}$.
For example, the commutation relations
of the
basic generators $L_m,\;J_m$ with some (of the spins 4, 5 and 6)
composite generators
($\sim L_n L_m,\; L_n J_m,\;J_n J_m $) contain in the r.h.s.,
apart from the composite generators, also basic generators which
appear just due to non-zero $c$ in (3.1). In what follows the presence of
such terms in the commutation relations will be very important
(cf. \cite{a12})
and this is the main reason why we should keep $c$ non-vanishing in (3.1)
\footnote{One can regard $c$ as a contraction parameter. After rescaling
$J_n  \equiv c^{-\frac{1}{2}} \tilde{J}_n,\; J_n^{(4)} \equiv c^{-1} \tilde
{J}_n^{(4)}$,  $c$ can be put equal to zero. In this limit (3.1) contracts
into the commutation relations of the ``classical $W_3$ algebra'' of ref.
\cite{a21}.}. As an example we quote the commutator
\be \label{1.20}
\left[ L_n, J_m^{(4)} \right] = (3n - m)\;J_{n+m}^{(4)} - \frac{4}{3}
(n^3 - n)\;L_{n+m}\;.
\ee
The second term in the r.h.s of (\ref{1.20}) is owing to the presence of
non-zero $c$ in (3.1).

Just as in \cite{a12}, while constructing a nonlinear realization of
$W_3^{\infty}$, we will deal not with the whole algebra, but with its
important subalgebra which contains all the spin $s$ generators($s\geq 2$)
with the indices varying from $- (s-1)$ to $\infty$
(this subalgebra is a genuine generalization of
the "truncated" Virasoro algebra, eq. (1.1)).
If one thinks of $W_3^{\infty}$ as an algebra of some $2D$ field variations
with holomorphic parameters (e.g., in the realization given in \cite{a12}),
the above subalgebra corresponds to restricting to the
parameters-functions regular at the origin. To simplify the terminology,
in what follows just this truncated algebra will generally be referred to as
$W_3^{\infty}$. We wish to point out that the higher-spin generators
of this algebra, when treated as composite, still involve the basic
generators with {\it all} conformal dimensions, both positive and negative.
For instance, in eq. (3.2) one restricts the index $n$ to vary in the range
$n\geq -3$, however the summation still goes over the whole range
$ -\infty < m < \infty $.

In the rest of this Section we list some subalgebras of the truncated
$W_3^{\infty}$ algebra which will be employed in our further
discussion. The proof of closeness of the relevant sets of generators
in most of the cases goes by a direct inspection and essentially relies upon
the
property that all
higher-spin generators in $W_3^{\infty}$ (with spins $\geq 4$) form an
invariant subalgebra \cite{a12}.

The reflection symmetry $n \rightarrow -n$ of the original relations (3.1)
implies the existence of a wedge subalgebra $W_{\wedge}$ in
$W_3^{\infty}$ \cite{a12}.
It is constituted by an infinite
number of generators, with the indices varying from $- (s-1)$
to $(s-1)$ for each spin $s$ {}\footnote{The precise relation of $W_3^{\infty}$
and this wedge algebra to the $W_{\infty}$-type algebras and their wedge
subalgebras (see, e.g. \cite{a22}) is not quite clear to us at present.
For instance, $W_{\infty}$ is known to contain each spin only once
while this is not true for the algebras $W_N^{\infty}$. It is
likely that $W_{\infty}$ can be obtained as a $N\rightarrow \infty$
limit and truncation of $W_N^{\infty}$.}
\be
W_{\wedge}= \left\{
\begin{array}{c}
L_{-1} \quad L_{0} \quad L_{1} \\
J_{-2} \quad J_{-1} \quad J_{0} \quad J_{1} \quad J_{2}  \\
J_{-3}^{(4)} \quad J_{-2}^{(4)} \quad J_{-1}^{(4)} \quad J_{0}^{(4)}
 \quad J_{1}^{(4)} \quad J_{2}^{(4)} \quad J_{3}^{(4)} \\
\ldots\ldots \ldots \ldots\ldots\ldots \ldots \ldots  \ldots \ldots \ldots  \\
\end{array}   \right\} \label{1.3}
\ee
(dots mean higher-spin generators with proper indices).
An interesting factor-algebra of this wedge algebra is the $sl(3,R)$ given
by:
\be \label{1.4}
W_{\wedge} /\{ J_{-3}^{(4)},\ldots ,J_{3}^{(4)},\ldots  \}
  \sim   sl(3,R) \quad .
\ee

One more important subalgebra of $W_3^{\infty}$ is constituted by the
following generators
\be  \label{1.5}
{\cal G} = \left\{ J_{-2}, J_{-1}, L_{-1}, L_0, J_0,
L_1, J_1, J_2, J_{n}^{(4)}\;\;(n\geq -3),
J^{(s)}_{n} \;\; (s\geq 5, \; n\geq -s +1 ) \right\}\;.
\ee
It is the maximal subalgebra of the truncated $W_3^{\infty}$. We see that it
is obtained
by adding to the wedge algebra (\ref{1.3}) an infinite rest of the
higher-spin generators with all positive conformal dimensions. All these
generators still form an ideal and the factor-algebra over this ideal
coincides with the $sl(3,R)$ (\ref{1.4}).

One may narrow the subalgebra (\ref{1.5}) successively removing from it
some generators.
It is easy to check that the sets of generators
\bea
{\cal H} &=& \left\{ J_{-1} + 2L_{-1}, L_0, J_0,
L_1, J_1, J_2, J_{n}^{(4)}\;(n\geq -3),
J^{(s)}_{n} \; (s\geq 5,  n\geq -s +1 ) \right\}  \label{1.6} \\
{\cal H}_{1} &=& \left\{ J_{-1} + 2L_{-1}, L_0, J_0,  J_{n}^{(4)}\;\;(n\geq
-3),
J^{(s)}_{n} \;\; (s\geq 5, \; n\geq -s +1 ) \right\}  \label{1.7} \\
{\cal H}_{2} &=& \left\{ J_{-1} + 2 L_{-1},  J_{n}^{(4)}\;\;(n\geq -3),
J^{(s)}_{n} \;\; (s\geq 5, \; n\geq -s +1 ) \right\}  \label{1.8}
\eea
still form subalgebras:
\be
{\cal H}_2 \subset {\cal H}_1 \subset {\cal H} \subset {\cal G}\;.  \nonumber
\ee
Note that ${\cal H}_1$ and ${\cal H}_2$ can be
extended by adding the generator $J_{-2}$
\be \label{1.10}
{\cal H}_{1}' =  {\cal H}_{1} \oplus J_{-2}\;,\;\;\;\; {\cal H}_{2}'
= {\cal H}_{2} \oplus J_{-2}\;.
\ee
The algebra ${\cal H}_{2}'$ has been already utilized
as the stability subgroup algebra in the Toda type nonlinear realization
of $W_3^{\infty}$ \cite{a12} (under the restriction to one
of two light-cone copies of $W_3^{\infty}$ considered in \cite{a12},
with the second $2D$ light-cone coordinate regarded as an extra ``time'').
The algebras  ${\cal H}$, ${\cal H}_1$ and  ${\cal H}_2$ will serve as
the stability subgroup algebras in new nonlinear realizations of
$W_3^{\infty}$ we will construct in the next Section.

\setcounter{equation}{0}
\section{Nonlinear realizations of $W_3^{\infty}$}

As has been argued in ref.\cite{a12},
extending the nonlinear realization method to the $W_N$ type symmetries
implies replacing
the latter by symmetries based on the linear algebras $W_N^{\infty}$ and
then constructing coset realizations of these $W_N^{\infty}$ symmetries
according to the general prescriptions of ref. \cite{a13}. Once this
is done, the original $W_N$ symmetry is expected to emerge as a
particular field realization of $W_N^{\infty}$. So a nonlinear (coset)
realization
of $W_3$ should always be understood as that of $W_3^{\infty}$. Just
this point of view has been put forward in ref.\cite{a12} and we will
pursue it here. In this way in \cite{a12} the $sl_3$ Toda
realization of $W_3$ symmetry has been reproduced starting
from a nonlinear realization of the product of two light-cone copies of
$W_3^{\infty}$ symmetries. Here we construct a set of nonlinear realizations
of one $W_3^{\infty}$. In Sect. 6 we will prove that they also amount to some
specific field realizations of the $W_3$ algebra (3.1), (3.2). In the next
Section we will show that these realizations bear a deep relation to
the Boussinesq equation and Miura maps for the latter.

Any nonlinear realization is quite specified by the choice of the
stability subgroup $H$ or, equivalently, its subalgebra ${\cal H}$.
So in the case at hand one should start by fixing the appropriate
${\cal H} \subset W_3^{\infty}$. Like in \cite{a12} we will always
place the entire set of higher-spin generators (starting with the spin $4$)
in the stability subalgebra and consider first a nonlinear realization
of $W_3^{\infty}$ with (3.7) as such a subalgebra. This choice can be
motivated by the following reasonings. In Sect. 2 we have learned that
the spin 2 conformal
stress-tensor $u(x)$ can be interpreted as the essential coset
field of the nonlinear realization of Virasoro symmetry $W_2$ with the
stability subgroup corresponding to the algebra (\ref{0.13}). There arises
a natural question whether it is possible to give an analogous coset
interpretation to both spin 2 and spin 3 currents associated with the $W_3$
algebra. Clearly, the appropriate stability subalgebra of $W_3^{\infty}$
should become (\ref{0.13}) upon reducing $W_3 \rightarrow W_2$, i.e.
taking away all generators except for the $W_2$ ones. The algebra (3.7) just
satisfies this criterion.

Inspecting the set of the $W_3^{\infty}$ generators which are out of (3.7),
i.e. belong to the coset, we find that the lowest dimension ones are
$J_{-2}$ $(cm^{-2})$, $L_{-1}$ $(cm^{-1})$, $L_{2}$ $(cm^{2})$
and $J_3$ $(cm^{3})$.
With the generator $L_{-1}$, like in the $W_2$ case, it is natural
to associate the coordinate $x$. The last two generators
have true dimensions for identifying the associate coset parameters,
respectively $u$ and $v$, with the spin 2 and spin 3 currents. All
higher-order coset generators have growing negative dimensions so the
corresponding parameters-fields are expected to be expressible in terms
of $u$ and $v$ by the inverse Higgs effect. But there still remains the
generator $J_{-2}$. The dimension of the coset parameter related to it
is inappropriate for treating this parameter as a field. On the other hand,
one cannot put $J_{-2}$ into the stability subgroup as its commutator with,
e.g., $J_1$ yields the coset generator $L_{-1}$ in the r.h.s. Thus, the only
possibility one may conceive is to treat this parameter
as an additional coordinate, we call it $t$,
in parallel with $x$ and to allow all coset
fields to depend on it. One observes that $t$ has the same dimension $cm^2$
as the evolution parameter of the Boussinesq equation \cite{a6}, so in what
follows we will
refer to it as to the ``time'' coordinate. Note that the interpretation of
$x$ and $t$ as the coordinates parametrizing the ``spatial'' and ``temporal''
directions is quite natural from the physical point of view, for the
translations along these directions are entirely independent as a
consequence of commutativity of the generators $L_{-1}$ and $J_{-2}$.

With all these remarks taken into account, an element of the coset space
we are considering can be parametrized as follows:
\be \label{2.1}
g = {\mbox e}^{tJ_{-2}}\; {\mbox e}^{xL_{-1}}\;
{\mbox e}^{\psi_3 L_3}\cdot \left( \prod_{n\geq 4}
{\mbox e}^{\psi_n L_n}\;{\mbox e}^{\xi_n J_n}\right) \cdot
{\mbox e}^{uL_{2}}
\;{\mbox e}^{vJ_{3}}
\ee

As usual in nonlinear realizations, the group G (associated with
$W_3^{\infty}$ in the present case) acts as left multiplications of the coset
element. This induces a group motion on the coset: the
coordinates $x$, $t$ together with the infinite tower of coset fields
$u(x,t)\;,\;v(x,t)$,
$\psi_{n}(x,t)\;,\;\xi_{n}(x,t)$ constitute a closed set under the group
action. We are postponing the discussion of all symmetries induced on
the coset parameters in this way to the Section 6 (their number is
infinite in accordance with the infinite dimensionality of $W_3^{\infty}$).

Besides the above minimal
coset we will need also extended cosets with the stability
subgroups generated by subalgebras ${\cal H}_{1}$ and ${\cal H}_{2}$ defined
in eqs. (\ref{1.7}), (\ref{1.8}). The relevant coset elements are represented
by
\be  \label{2.2}
g_1 = g {\mbox e}^{u_1 L_1} {\mbox e}^{v_1 J_1} {\mbox e}^{v_2 J_2}
\ee
\be
g_2 = g_1 {\mbox e}^{u_0 L_0} {\mbox e}^{v_0 J_0}\;, \label{2.3}
\ee
where $u_1,\;v_1,\;v_2,\;u_0,\;v_0$ are additional parameters-fields, all
given on the space $\left\{ x,\;t \right\}$. It is worth mentioning that
the subalgebra (\ref{1.7}) and the associated nonlinear realization of
$W_3^{\infty}$ generalize the subalgebra (\ref{0.10}) and  the
realization of $W_2$ related to this choice. In the realization of
$W_3^{\infty}$ on elements (\ref{2.3}) all Virasoro generators
belong to the coset, so it is an extension of the realization (\ref{0.4}).

Let us now turn to constructing Cartan forms for the above cosets.
Like in the $W_2$ case these are defined by the generic relation
\be \label{2.4}
\Omega \equiv g^{-1}\;dg = \sum_{n\geq -1} \omega_n L_n + \sum_{n\geq -2}
\theta_n J_n + {\mbox{Higher-spin contributions}}\;,
\ee
and by the analogous ones for two other cases, with $g$ replaced by $g_1$ or
$g_2$.

The explicit expressions for the forms which we actually need
can be obtained by using solely the commutation relations (3.1) and
(\ref{1.20}). For the realization (\ref{2.1}) these are
\bea
 \omega_0 &=& 0\;, \;\omega_{1} = 160 v dt - 3u dx \nonumber \\
 \theta_{-1} &=& 0\;,\; \theta_{0} = -6udt\;,\;
\theta_{1} = -8 \psi_3 dt\;,\;
\theta_2 = 12 u^2 dt -( 5v dx + 10 \psi_4 dt)
\label{2.5}
\eea
\bea
\omega_{-1} &=& dx \nonumber \\
\omega_2 &=& du + 320 \xi_4 dt - 4\psi_3 dx \nonumber \\
\omega_3 &=&
d\psi_3 + ( \frac{3}{2} u^2 - 5\psi_4 ) dx + (560 \xi_5 - 240 u v) dt
\nonumber \\
\omega_4 &=& d\psi_4 - 6( \psi_5 - u \psi_3) dx + ( 896\xi_6 - 320 v \psi_3 +
6384 u \xi_4)dt  \label{2.6}
\eea
\bea
\theta_{-2} &=& dt \nonumber \\
\theta_3 &=& dv - 6\xi_4 dx + (36 u \psi_3 - 12\psi_5) dt \nonumber \\
\theta_4 &=&
d\xi_4 - (7\xi_5 - 2vu) dx + (20 \psi_3^3 + 48 u \psi_4 + 80 v^2 + 8u^3
- 14 \psi_6 ) dt\;. \label{2.7}
\eea
The Cartan forms (\ref{2.5}) belong to the stability subalgebra and transform
inhomogeneously under
left
$W_3^{\infty}$ shifts. The forms (\ref{2.6}), (\ref{2.7}) are
associated with the coset generators, so they are transformed into themselves
and other coset forms.

In the realization corresponding to the coset (\ref{2.2}) some Cartan
forms on the stability subalgebra get contributions from the new coset
fields $u_1,\;v_1,\;v_2$ and, besides, the Cartan forms
$\omega_1,\;\theta_1,\; \theta_2$ become belonging to the coset. Due to the
special arrangement of factors in (\ref{2.2}) the forms associated with
other coset generators are linear combinations of the previous coset forms
with the coefficients depending on the new coset fields.
We give here the explicit expressions only for the new coset forms
\bea
\omega_1 &=& du_1 + (u_1^2 + 12 v_1^2 - 3u) dx +
(160 v + 48 u v_1 - 48 v_1 u_1^2 -
64 u_1 v_2 + 64 v_1^3) dt \label{2.8} \\
\theta_1 &=& dv_1 + (2u_1 v_1 - 4v_2) dx +
(12u u_1 - 8\psi_3 - 4u_1^3 -64 v_1v_2 - 16u_1 v_1^2)
dt  \label{2.9} \\
\theta_2 &=& dv_2 + du_1 v_1 + (4v_1^3 + 4u_1 v_2 +
u_1^2v_1 -3uv_1 -5v) dx + (12u^2 -10 \psi_4 +u_1^4     \nonumber \\
	 & &   + 16v_1^4-64v_2^2 -64 u_1v_1v_2 -24u_1^2v_1^2
-6 uu_1^2 +24uv_1^2 +160 vv_1 +8u_1\psi_3).\label{2.10}
\eea

Finally, the net effect of passing to the coset (\ref{2.3}) is
homogeneous rotations of the previously defined forms by the group factors
with generators $L_0$ and $J_0$ and extension of the set of coset forms
by $\omega_0$ and $\theta_0$. The latter forms are given by
\bea
\omega_0 &=& du_0 + (32 u_1 v_1 + 64v_2) dt - 2u_1 dx \nonumber \\
\theta_0 &=& dv_0 + (6u_1^2 - 24 v_1^2 -6u) dt -3 v_1 dx \;. \label{2.11}
\eea

As the last remark we mention that the form $\Omega$ (\ref{2.4}) (and its
analogs for the
cosets (\ref{2.2}) and (\ref{2.3}) $\Omega_1$ and $\Omega_2$) by
definition satisfies the Maurer-Cartan equation
\be \label{2.12}
d^{ext}\Omega = \Omega \wedge \Omega \;.
\ee

\setcounter{equation}{0}
\section{Boussinesq equations and Miura maps from covariant
reduction of the cosets of $W_3^{\infty}$}

Here we generalize to the $W_3^{\infty}$ cosets the inverse Higgs procedure
(alias covariant reduction)
which has been already applied in
Sect.2 to simpler examples of nonlinear realizations
of Virasoro symmetry. A new feature of this procedure in the case at hand
is that it will lead not only to the kinematic equations for expressing
higher-order coset fields in terms of a few essential ones but also to the
dynamical equations for the latter. This is directly related to the presence
of the extra time coordinate $t$.

We begin with the coset (\ref{2.1}). A natural generalization of the
constraints (\ref{0.14}) is as follows
\be \label{3.1}
\omega_n = 0\;, \;\;\forall_{n\geq 2} \;\;\;\;\theta_m = 0\;,\;\;
\forall_{m\geq 3}\;.
\ee
Upon imposing these constraints, the one-form $\Omega$ (\ref{2.4}) defined
originally on
the entire algebra $W_3^{\infty}$ is reduced to the one-form valued
in the algebra ${\cal G}$ (\ref{1.5})
\be \label{3.2}
\Omega \Rightarrow \Omega^{red} \subset {\cal G}\;.
\ee
In accordance with the terminology explained in Sect.2, ${\cal G}$
is the covariant reduction subalgebra in the present case. Taking into
account that ${\cal H}$ coincides with the $sl(3,R)$ (\ref{1.4}) modulo
higher-spin generators, one may, without loss of generality, regard
just this $sl(3,R)$ as the reduction subalgebra and consider only
the $sl(3,R)$ part of $\Omega^{red}$. This part obeys
the Maurer-Cartan equation (\ref{2.12})
in its own right, without any contributions
from the higher-spin generators. We will make use of this observation a bit
later.

Let us now inspect eqs. (\ref{3.1}). As opposed to the $W_2$ constraints
(\ref{0.14}), each of eqs. (\ref{3.1}) actually produces two equations, for
the coefficients of the differentials $dx$ and $dt$. Using the explicit
structure of the lowest coset forms, eqs. (\ref{2.6}) and (\ref{2.7}), one
finds
\bea
\psi_3 &=& \frac{1}{4} u' \nonumber \\
\psi_4 &=& \frac{1}{5} ( \frac{3}{2} u^2 + \frac{1}{4} u'') \nonumber \\
\psi_5 &=& \frac{1}{60} ( 21 u u' + \frac{1}{2} u''') = \frac{1}{12} (\dot{v}
+ 9u\dot{u}) \label{3.3}
\eea
\begin{equation}
\xi_4 = -\frac{1}{320} \dot{u} = \frac{1}{6} v' \;\; \mbox{etc.} \label{3.4}
\end{equation}
We see that the higher-order coset fields are
expressed by the inverse Higgs constraints (\ref{3.1}) in terms of two
independent ones, $u(x,t)$ and $v(x,t)$, thus generalizing an analogous
phenomenon of the $W_2$ case. However, for all coset fields, except
$\psi_3,\psi_4 $
there simultaneously appear two expressions coming from equating
to zero the
coefficients of the differentials $dx$ and $dt$ in the appropriate forms.
Requiring these expressions to be compatible amounts to the set of dynamical
equations
\bea
\dot{u} &=& -\frac{160}{3} v' \nonumber \\
\dot{v} &=& \frac{1}{10}u''' - \frac{24}{5} u'u\;, \label{3.5}
\eea
which is recognized as the Boussinesq equation \cite{a6} (after appropriate
rescalings). Its another, second-order form is obtained by differentiating
the first equation in (\ref{3.5}) with respect to $t$ and then using the
second equation
\be \label{3.6}
\ddot{u} = -\frac{16}{3} u'''' + 128 (u^2)''\;.
\ee
With making use of the Maurer-Cartan equation (\ref{2.12}) one may show that
the rest of
constraints (\ref{3.1}) does not imply any further dynamical restrictions on
the fields $u,\;v$ and serves only for the covariant elimination of
higher-order coset fields.

Thus we have succeeded in deducing the Boussinesq equation from a nonlinear
realization of $W_3^{\infty}$ like it has been done in \cite{a12} for
the $sl_3$ Toda equations (starting with a nonlinear realization
of two copies of $W_3^{\infty}$). This shows a close relation of the
Boussinesq equation to the intrinsic geometry of $W_3^{\infty}$: it reveals
a nice geometric
meaning as one of the constraints singling out a finite-dimensional
geodesic hypersurface in the coset of $W_3^{\infty}$ over the subgroup
with the algebra ${\cal H}$ (\ref{1.8}). This hypersurface is homeomorphic
to the two-dimensional coset of the group with the algebra ${\cal G}$
(\ref{1.5}) over the
subgroup with the algebra ${\cal H}$ (\ref{1.8}). Taking account of the
fact that the higher-spin generators drop out after such a
factorization, this coset coincides with that of the group $SL(3,R)$
with the generators
(\ref{1.4}) over its six-parameter
Borel subgroup generated by $\left\{J_{-1} + 2L_{-1},\;
J_0,\;L_0,\;L_1,\;J_1,\;J_2 \right\}$. The coordinates $x$ and $t$
parametrize this coset while the fields $u$ and $v$ describe the embedding
of it as a hypersurface in the original coset space of $W_3^{\infty}$.

The Boussinesq equation is known to be completely integrable: it
possesses a zero-curvature representation and the related Lax pair
on the algebra $sl(3,R)$ \cite{{a9},{a18}}. It is instructive to see how these
integrability properties are reproduced in the present geometric picture.
After substitution of the expressions for higher coset fields, the most
essential, $sl(3,R)$ part of the one-form $\Omega^{red}$ defined in eq.
(\ref{3.2}) reads
\be \label{3.7}
\Omega^{red} = ( L_{-1} - 5v J_2 - 3u L_1) dx +
[ 160 v L_1 + (9u^2 - \frac{1}{2}
u'') J_2 + J_{-2} - 6u J_0 - 2u'J_1 ] dt\;.
\ee
As has been mentioned before, the original Maurer-Cartan equations for this
one-form are closed modulo higher-spin generators. Discarding the higher-spin
pieces in the commutators of the $sl(3,R)$ generators in $\Omega^{red}$,
one easily establishes that the Maurer-Cartan equation
\be \label{3.8}
d^{ext} \Omega^{red} = \Omega^{red} \wedge \Omega^{red}
\ee
implies the Boussinesq equation (\ref{3.5}) and so provides the
zero-curvature representation for the latter. Recall that the original
Maurer-Cartan equation (\ref{2.12}) was purely kinematical. It becomes
dynamical after invoking the covariant reduction constraints (\ref{3.1})
(\ref{3.2}). It should be emphasized that just these constraints are
primary dynamical restrictions on the fields $u$ and $v$ in the present
approach; the zero-curvature representation (\ref{3.8}) is their consequence.
This feature is typical for all other examples where the
covariant reduction proved to be efficient \cite{{a14},{a15},{a12}}.

To obtain a Lax representation from eqs. (\ref{3.7}), (\ref{3.8}), one
introduces
the ``covariant derivatives''
$$ \frac{\partial}{\partial t} + A_{t}\;,\;\;\; \frac{\partial}{\partial x}
+ A_x\;,$$
where the $sl(3.R)$ algebra valued connections $A_t$ and $A_x$ coincide with
the coefficients of $dt$ and $dx$ in (\ref{3.7}), and rewrites eq. (\ref{3.8})
as the condition of commutativity of these covariant derivatives. Note that
in this way one obtains just the Drinfel'd-Sokolov type Lax pair \cite{a9}
for the
Boussinesq equation (after choosing $sl(3,R)$ generators in the
fundamental $3 \times 3$ matrix representation).

Let us turn to discussing the coset (\ref{2.2}). As was already mentioned,
it is an extension of the $W_2$ coset associated with the stability
subalgebra (\ref{0.10}). So the relevant set of the covariant reduction
constraints should be an appropriate generalization of the set (\ref{0.11}):
\be  \label{3.9}
\omega_n = 0\;,\;\; \forall_{n\geq 1}\;, \;\;\;\; \theta_m = 0\;, \;\;
\forall_{m\geq 1}\;.
\ee
It includes the previous set (\ref{3.1}) and, in addition, implies vanishing
of the Cartan forms $\omega_1,\;\theta_1,\;\theta_2$. The fields $v$ and $u$
still obey the Boussinesq equation (\ref{3.5}) but now they are expressed
(like $v_2$) through the fields $v_1$ and $u_1$ which are
the only independent coset fields for the realization at hand. Bearing in
mind the explicit expressions for the additional coset forms (eqs. (\ref{2.8})
- (\ref{2.10})), one finds:
\bea
u &=& \frac{1}{3} ( u_1' + u_1^2 + 12 v_1^2) \label{3.10} \\
v &=& \frac{1}{5} (\frac{1}{4} v_1'' + \frac{1}{2} u_1'v_1 + \frac{3}{2}
u_1 v_1' + 2u_1^2v_1 - 8v_1^3 )\;. \label{3.11}
\eea
By the same mechanism as in the previous case (compatibility between the
equations coming from the coefficients of $dx$ and $dt$ in the
appropriate forms) one also obtains the dynamical restrictions on
the fields $u_1$ and $v_1$
\bea
\dot{u}_1 &=& -8 (v_1' + 4 u_1v_1)' \nonumber \\
\dot{v}_1 &=& \frac{2}{3}( u_1' - 2u_1^2 + 24 v_1^2)'\;. \label{3.12}
\eea
These equations can be easily checked to be consistent with the Boussinesq
equation: differentiating (\ref{3.10}), (\ref{3.11}) with respect to
$t$ and making use of eqs. (\ref{3.12}) one obtains just (\ref{3.5}).

The expressions (\ref{3.10}), (\ref{3.11}) are a genuine generalization
of eq. (2.13) and provide a Miura map of the $W_3$ currents $u$ and
$v$ onto the two independent $U(1)$ Kac-Moody currents $u_1$ and $v_1$. Thus
in the present case this map also gets a geometric interpretation as the
covariant relations between the fields parametrizing the coset of
$W_3^{\infty}$ symmetry.

By analogy with the modified KdV equation, it is
natural to call eqs. (\ref{3.12}) the modified Boussinesq equation. It can be
rederived from the vanishing of the curvature of the
reduced Cartan form $\Omega_1^{red}$ (with the higher-spin generators
factored out)
\bea
\Omega_1^{red} &=&  [ J_{-2} - 4u_1J_{-1} +
 16( v_1' + 4u_1v_1) L_0 + 2 (2u_1^2 -24v_1^2 -u_1') J_0 +
16v_1L_{-1}] dt  \nonumber \\
   && + (L_{-1} - 2u_1 L_0 -3v_1J_0) dx  \label{3.13}
\eea
and so is integrable like the Boussinesq equation. One observes that
$\Omega_1$ is given on the five-dimensional subalgebra of the $sl(3,R)$.
So in the present case the covariant reduction actually leaves us with
the coset space of the group associated to this subalgebra over the subgroup
generated by $J_{-1},\;J_0$ and $L_0$. Once again, the coordinates $t$ and
$x$ are the parameters of this coset while the fields $u_1$ and $v_1$
specify how the latter is embedded into the original $W_3^{\infty}$ coset.

Finally, let us see which new features are brought about by passing to the
coset $g_2$ defined in eq.(\ref{2.3}). In this case the essential coset fields
are $u_0$, $v_0$ and, in addition to the previous
constraints, one should require vanishing of the two newly appearing coset
Cartan forms $\omega_0$ and $\theta_0$ given by eq. (\ref{2.11})
\bea
\omega_0 &=& \theta_0\; =\; 0\;, \Rightarrow  \label{3.14} \\
u_1 &=& \frac{1}{2}u_0'\;,\;\; v_1 \;=\; \frac{1}{3}v_0'\;, \label{3.15} \\
\dot{u}_0 &=& -\frac{16}{3} ( v_0'' + 2 u_0'v_0')\;,\;\; \dot{v}_0 = u_0'' -
(u_0')^2 + \frac{16}{3} (v_0')^2 \;. \label{3.16}
\eea
The covariant relations (\ref{3.15}) are analogs of the first of eqs.
(\ref{0.9}), they give a further Miura
transformation from the $U(1)$ Kac-Moody currents
$u_1$, $v_1$ and the
$W_3$ currents $u$, $v$ to the scalar fields $u_0$ and $v_0$. For $u$ and $v$
one obtains the representation
\bea
u &=& \frac{1}{6}\; [\; u_0'' + \frac{1}{2} (u_0')^2 + \frac{8}{3} (v_0')^2\; ]
\nonumber \\
v &=& \frac{1}{5}\; [\; \frac{1}{12} v_0''' +
\frac{1}{12}u_0''v_0' + \frac{1}{4}
u_0'v_0'' + \frac{1}{6} (u_0')^2 v_0' - \frac{8}{27} (v_0')^3\; ] \;,
\label{3.17}
\eea
which, after appropriate rescalings, is recognized as the free-field
Feigin-Fuchs type representation for the $W_3$ currents \cite{{a4},{a10}}.
Once again, the dynamical equation (\ref{3.16}) induces for the sets
$u$, $v$ and
$u_1$, $v_1$
the Boussinesq equation (\ref{3.5}) and the modified Boussinesq
equation (\ref{3.12}). It amounts to the zero-curvature representation for the
reduced one-form
\be
\Omega_2^{red} = \omega_{-1}L_{-1} + \theta_{-1}J_{-1} + \theta_{-2}J_{-2}
\label{3.18}
\ee
with
\bea
\omega_{-1} &=& {\mbox e}^{-u_{0}} [\; 4u_0'\;
{\mbox {sinh}} (4v_0) dt + (dx + \frac{16}{3}
v_0'dt)\; {\mbox {cosh}} (4v_0)\;] \nonumber \\
\theta_{-1} &=& -{\mbox e}^{-u_{0}} [\; 2 u_0'\;
{\mbox {cosh}} (4v_0)dt + \frac{1}{2}
(dx + \frac{16}{3}v_0'dt)\; {\mbox{sinh}}(4v_0)\; ] \nonumber \\
\theta_{-2} &=& {\mbox e}^{-2u_0} dt \;.
\label{3.19}
\eea
Thus in the present case the original coset of $W_3^{\infty}$ has been
covariantly
reduced to the two-dimensional
coset of the three-parameter subgroup with the generators
$L_{-1},\;J_{-1},\;J_{-2}$ over the one-parameter subgroup generated by
$J_{-1}+2L_{-1}$.

In conclusion of this Section we briefly discuss the relation to the
Hamiltonian approach which provides one more link between the
Boussinesq equation and the algebra $W_3$. It is known \cite{a6}
that this equation can be interpreted as a Hamiltonian flow on
$W_3$. Namely, it possesses the second Hamiltonian structure
with the
Poisson brackets between $u$ and $v$ forming  $W_3$
\bea
\dot{u} &=& \{ u,\;H \}\;, \dot{v}\; =\; \{ v,\;H \}\;, \label{3.20} \\
H &=& \frac{40c}{3}\int dx v(x,t)  \label{3.21} \\
\{ u(x,t),\;u(y,t) \} &=&\frac{2}{c}\left[ \frac{1}{6}\;
\frac{\partial^3}{\partial y^3}
+2u\; \frac{\partial}{\partial y}-u'\right]\delta (x-y)\nonumber \\
\{ u(x,t),\;v(y,t) \} &=& -\frac{2}{c}\left[ 3v \; \frac{\partial}{\partial y}+
v' \right] \delta (x-y)\nonumber \\
\{ v(x,t),\;v(y,t) \} &=&\frac{3}{10c}\left[
-\frac{1}{48}\; \frac{\partial^5}{\partial y^5}
+\frac{5}{4}u\;\frac{\partial^3}{\partial y^3}
+\frac{15}{8}u'\; \frac{\partial^2}{\partial y^2}- \right. \nonumber \\
 && \left.-\left( -\frac{9}{8}u''+12u^2\right)\frac{\partial}{\partial y}
 -\left( -\frac{1}{4}u'''+12uu' \right) \right] \delta(x-y)
\label{3.22}
\eea
where the fields in the r.h.s. are evaluated at the point $y$.
Decomposing $u$ and $v$ in the Fourier modes
with respect to $x$, one observes that the algebra (\ref{3.22})
implies for these modes just the $W_3$ algebra relations (3.1), (3.2).

We wish to point
out that this Hamiltonian formalism matches very naturally with our
nonlinear realization approach, though the precise relation between these two
is as yet not clear to us. If one substitutes the Fourier
decomposition of $v$ in the Hamiltonian (\ref{3.21}) and integrates over $x$,
$H$ is recognized, up to a scale factor, as the generator $J_{-2}$, just the
time translation generator in the nonlinear realization scheme.

As far as the modified Boussinesq equations (\ref{3.12}), (\ref{3.16}),
are concerned they
can be given the standard Hamiltonian form like (\ref{3.20}) with the same
Hamiltonian (\ref{3.21}) expressed in terms of $u_1,\;v_1$ or $u_0,\;v_0$ by
eqs. (\ref{3.11}), (\ref{3.17}) and with the following underlying Poisson
structure
\begin{equation}
\{ u_1 (x,t),u_1(y,t) \} =\frac{3}{c}\; \frac{\partial}{\partial y}
\delta (x-y)\;,\;
\{ v_1(x,t),v_1(y,t) \}=
\frac{1}{4c}\; \frac{\partial}{\partial y}\delta (x-y) , \label{3.23}
\end{equation}
$$\{ u_1(x,t),v_1(y,t) \} = 0, $$
\begin{equation}
\{ u_0 (x,t),u'_0(y,t) \}= -\frac{12}{c}\; \delta (x-y)  \;,\;
\{ v_0(x,t),v'_0(y,t) \}= -\frac{9}{4c}\; \delta (x-y) , \label{3.24}
\end{equation}
$$\{ u_0(x,t),v_0(y,t) \} = 0\;. $$
These Poisson brackets are characteristic of the $U(1)$ Kac-Moody currents and
free scalar fields.

\setcounter{equation}{0}
\section{$W_3$ symmetry of Boussinesq equations as left $W_3^{\infty}$
shifts}

When studying integrable systems, one of the most important questions
is which symmetries preserve the given equation. In the previous Sections we
reformulated the
Boussinesq equation in the framework of the nonlinear realizations approach as
one of the covariant conditions which single out a two dimensional geodesic
hypersurface in the coset (\ref{2.1})
of ${W^\infty_3}$. Symmetries of Boussinesq equation are then the set
of ${W^\infty_3}$ transformations acting on the coset elements (\ref{2.1})
from the left
\begin{equation}\label{q1}
g_{0}(\lambda) \; g(x,t,u,v,\ldots )=
g({\tilde x},{\tilde t},{\tilde u},{\tilde v},\ldots )
\; h(\lambda ,x,t,u,v,\ldots )\;.
\end{equation}
Here $g_{0}(\lambda )$ is an arbitrary element of ${W^\infty_3}$ with
constant parameters and the induced element
$h$ belongs to the stability subgroup $H$ generated by the set of generators
(\ref{1.6}).

In principle one could directly evaluate ${\tilde x},{\tilde t},{\tilde u},
{\tilde v}$ by using eq.(\ref{q1}) and the commutation relations (3.1).
However, in the case at hand even the infinitesimal transformations of the
fields $\delta u,\delta v$ and coordinates $\delta x, \delta t$ are very
complicated functions of time $t$
and all the higher-order coset fields $\psi_n,\xi_m$, so it is not too
enlightening to try to give them explicitly. Below we
will pursue another approach, in which transformation properties of the
fields $u,v$ and coordinates $x,t$
are obtained together with an additional condition fixing
the $t$ dependence of the transformation parameters.

Let us begin by writing down the Cartan forms for the transformed coset
(\ref{q1})
\bea
\Omega &=&   h^{-1} \tilde{\Omega} h  + h^{-1}dh   \label{q2} \\
\tilde{\Omega} &\equiv &
g^{-1}({\tilde x},{\tilde t},{\tilde u},{\tilde v},\ldots )
\;d g({\tilde x},{\tilde t},{\tilde u},{\tilde v},\ldots )\;,    \label{q2.0}
\eea
where we have made use of the fact that the group parameters in (\ref{q1})
do not depend on $x,\;t$.
Now, the induced
element $h$ of the stability subgroup can be parametrized as follows
\begin{equation}\label{q3}
h= {\mbox e}^{a_0 L_0}\;{\mbox e}^{a_1 L_1}\;{\mbox e}^{b_{-1}J_{-1}}\;
{\mbox e}^{b_{0}J_{0}}\;{\mbox e}^{b_{1}J_{1}}\;
{\mbox e}^{b_{2}J_{2}} \; \tilde{h}\; ,
\end{equation}
where ${\tilde h}$ stands for the factors spanned by higher-spin
generators. Keeping in mind
that the higher-spin generators form an ideal in the stability subalgebra
and  comparing the Cartan forms associated with the generators
$L_{-1},L_0,L_1,J_{-2},J_{-1},J_0,J_1,J_2$ in both sides of
(\ref{q2}) we immediately obtain the following set of equations
(let us remind that the parameters $a$ and $b$ are infinitesimal)
\begin{eqnarray}\label{q4}
\omega_{-1} & = & \tilde{\omega}_{-1} -a_0 \tilde{\omega}_{-1} +
 16b_1\tilde{\theta}_{-2}+8b_{-1}\tilde{\theta}_{0} \nonumber \\
 0 & = & da_{0}+64b_{2}\tilde{\theta}_{-2} +8b_{-1} \tilde{\theta}_{1} -
 2a_1\tilde{\omega}_{-1} \nonumber \\
\omega_{1} & = & da-1+\tilde{\omega}_{1} +a_0 \tilde{\omega}_{1} -
 8b_1\tilde{\theta}_{0}-8b_{0}\tilde{\theta}_{1}
  -16b_{-1}\tilde{\theta}_{2}\nonumber \\
\theta_{-2} & = & \tilde{\theta}_{-2} -2a_0 \tilde{\theta}_{-2} -
 b_{-1}\tilde{\omega}_{-1}  \\
0 & = & db_{-1}-4a_1 \tilde{\theta}_{-2} -
 2b_0\tilde{\omega}_{-1} \nonumber \\
\theta_{0} & = & db_0+\tilde{\theta}_{0} +3b_{-1} \tilde{\omega}_{1} -
 3b_1\tilde{\omega}_{-1} \nonumber \\
\theta_{1} & = & db_1+\tilde{\theta}_{1} +a_0 \tilde{\theta}_{1} -
 2a_1\tilde{\theta}_{0}-4b_{2}\tilde{\omega}_{-1}
+2b_{0}\tilde{\omega}_{1} \nonumber \\
\theta_{2} & = & db_2+\tilde{\theta}_{2} +2a_0 \tilde{\theta}_{2} -
 a_1\tilde{\theta}_{1}+b_{1}\tilde{\omega}_{1} \quad .\nonumber \\
\end{eqnarray}

{}From the explicit expressions for the lowest Cartan forms
(\ref{2.5}), (\ref{2.6}) and (\ref{2.7}) we obtain
the following equations for the variations of fields and coordinates:
\begin{eqnarray}
\delta u &\equiv& \tilde{u}(x + \delta x,\;t + \delta t) - u(x,\;t)\; = \;
\frac{1}{6} \left[
\frac{1}{2} \dot{\left( \delta t \right)}'' - 6 u \dot{( \delta t)}+
 480 v \left( \delta t \right)' \right] \nonumber \\
\delta v &\equiv& \tilde{v}(x + \delta x,\;t + \delta t)-v(x,\;t)  \nonumber \\
 &=& -\frac{1}{160} \left[
\frac{1}{4} \ddot{\left( \delta t \right)}' - 6 u \dot{( \delta x)}+
 240 v \dot{(\delta t)} -8u'\left( \delta t \right)''
+8u''\left( \delta t \right)'
 \right] \label{q5} \\
\dot{(\delta t)} & = & 2 \left( \delta x \right)' \nonumber \\
\dot{(\delta x)} & = & 16 \left[ 8u \left( \delta t \right)'
 -\frac{1}{6} \left( \delta t \right)''' \right] \label{q6}
\end{eqnarray}

Several comments are needed concerning the transformations properties
(\ref{q5}), (\ref{q6}).

First of all, the time dependence of variations of
the coordinates
$\delta x,\;\delta t$ is controlled by the differential equations (\ref{q6})
which involve the field $u(x,t)$. It is hardly possible
to find the general solution of these equations in a closed explicit form.
Nonetheless, after
expanding
the coordinate variations and the field $u$ in Taylor series with respect
to $t$
\be \label{q7}
\delta t =  \sum_{n=0}^{\infty} \delta_n t(x)\;t^n\;,\;\;
\delta x  = \sum_{n=0}^{\infty} \delta_n x(x)\;t^n \;,\;\;
u(x,t)  =  \sum_{n=0}^{\infty} \frac{\partial^n u}{\partial t^n}\;t^n\;,
\ee
the role of eqs.(\ref{q6}) is reduced to expressing all functions
$\delta_n t(x)\;,\;\delta_m x(x)$
through two independent functions $\delta_0 t(x)$ and $\delta_0 x(x)$.
So the transformations (\ref{q5}) are actually specified by the two
functions of the coordinate $x$, much like the realization of one of the
light-cone $W_3$'s
in the $sl_3$ Toda system \cite{a12}. Thus we have proven that the
nonlinear realization of $W_3^{\infty}$ in the coset considered is
reduced to a kind of $W_3$ transformations of the fields $u$ and $v$.
The same of course is true for other nonlinear realizations, with $u_1,\;v_1$
and $u_0,\;v_0$ as the essential coset parameters.

Secondly, after passing to the active form of the transformations of $u,v$
$$
\tilde{\delta}u = \delta u - \delta t\; \dot{u}  - \delta x\; u' \quad , \quad
\tilde{\delta}v = \delta v - \delta t\; \dot{v}  - \delta x\; v'
$$
and eliminating time derivatives by the Boussinesq equation
(\ref{3.5}) and
constraints (\ref{q6}) we obtain the standard $W_3$ transformations for
the spin 2 and spin 3 currents:
$$ u \equiv -T \quad , \quad v \equiv -\frac{3}{80}
J \quad , \quad \delta x \equiv -f
\quad , \delta t \equiv g$$
\begin{eqnarray}
\tilde{\delta}_{conf} T & = & \frac{1}{6}f''' +2 f'T+f T' \nonumber \\
\tilde{\delta}_{conf} J & = & 3 f'J+f J' \label{q8} \\
\tilde{\delta}_{w} T & = & 3g'J+2g J' \nonumber \\
\tilde{\delta}_{w} J & = & -\frac{2}{9}g''''' -\frac{40}{3}g'''T-20g''T'
-12g'T''+32g'\Lambda -\frac{8}{3}gT'''+16g\Lambda'' \label{q9}
\end{eqnarray}
where
$$\Lambda = -4 T^2 \quad . $$
and parameters are still subject to the constraints (\ref{q6}).

These transformations and constraints are just those deduced in a
recent paper \cite{a18} starting with a Lax representation for the Boussinesq
equation. In our scheme they come out in a nice geometric way as the
$W_3^{\infty}$ group motions on the set of essential coset parameters
$\{ x,\;t,\;u(x,\;t),\;v(x,\;t) \}$. Invariance of the Boussinesq equation
under these transformations does not need to be checked, it directly
stems from the fact that this equation is a dynamical part of the inverse
Higgs constraints (\ref{3.1}) which are $W_3^{\infty}$-covariant by
construction.

Before passing to concluding remarks let us comment on the cosets of
$W_3^{\infty}$ corresponding to the choice of ${\cal H}_1'$, ${\cal H}_2'$
(eq.(\ref{1.10})) as the stability subgroup algebras. In these cases the
generator $J_{-2}$ belongs to the stability subgroup, so one is left
with one coordinate $x$, on which all other coset parameters are
assumed to depend. The covariant reduction constraints, like in
nonlinear realizations of Virasoro symmetry (Sect. 2), do not produce
any dynamical restrictions and serve entirely for the covariant elimination
of the higher-dimension coset fields via the essential ones ($u_1(x),\;v_1(x)$
or $u_0(x),\;v_0(x)$). In particular, the Miura maps (\ref{3.10}),
(\ref{3.11}),
(\ref{3.15}) and (\ref{3.16}) arise as before. For the independent coset fields
the left $W_3^{\infty}$ shifts generate the standard $sl_3$ Toda-type $W_3$
transformations parametrized by two functions of $x$ which collect
the constant parameters associated with the spin 2 and spin 3
generators $L_n$ and $J_n$. These transformations are just those deduced in
\cite{a12} (as far as one light-cone copy of $W_3^{\infty}$ is considered).

These realizations actually bear a tight relation to those associated with
the
cosets (\ref{2.2}), (\ref{2.3}). One may pass to a different parametrization
of
these coset elements where the time factor ${\mbox e}^{\tilde{t}J_{-2}}$
stands from
the right ($\tilde{t}$ is related to $t$ via a complicated field-dependent
redefinition). Then one may check that the first-order time derivatives of
the coset
fields in this new parametrization are transformed into themselves
under left $W_3^{\infty}$ shifts and so can be self-consistently put equal
to zero,
thus eliminating any time dependence of the coset fields. This is
equivalent to placing $J_{-2}$ from the beginning in the stability
group algebra.

\setcounter{equation}{0}
\section{Conclusion}

In this paper we have revealed a new kind of the relationship between
$W_3$ symmetry and Boussinesq as well as modified Boussinesq equations:
these have been found
to emerge in a geometric way as covariant dynamical constraints on the
parameters of
some coset manifolds of $W_3^{\infty}$ symmetry associated with $W_3$. The
Miura maps relating these equations to each other arise as a sort of
covariant kinematical constraints on the coset parameters. Put together,
these
constraints can be interpreted as the conditions singling out
finite-dimensional
geodesic hypersurfaces in the original infinite-dimensional coset
manifolds. The spin 2 and spin 3 $W_3$ currents and the introduced via
Miura maps two spin 1 $U(1)$ Kac-Moody currents and two spin 0 scalar fields
come out as the essential parameters of three  coset manifolds of
$W_3^{\infty}$ embedded into each other. Thus the considered Boussinesq-type
equations, related Miura maps, involved currents and fields prove
to be intimately linked to the intrinsic geometries of the coset manifolds
of $W_3^{\infty}$, just like the $sl_3$ Toda equations \cite{a12}. The
common geometric origin of the latter equations and the Boussinesq ones
suggests a deep connection between them which can hopefully be exposed
most clearly within the present approach. The understanding of this
relationship could have important implications, e.g. in $W_3$ strings and $W_3$
gravity.

An interpretation of Miura maps as the covariant relations between the
fields parametrizing coset manifolds of the $W_N^{\infty}$ type symmetries
seems to be especially useful in searching for free-field representations
of the currents generating more complicated $W$ symmetries and their
superextensions. Usually this is a subject of some guess-work.
As we have argued in \cite{a12} and this paper, within the present approach
finding such
representations becomes more straightforward and algorithmic. One starts by
defining the appropriate linear $W_3^{\infty}$ type symmetry and its cosets,
then construct the Cartan forms and finally impose suitable
covariant reduction constraints. Doing so, we have recently found, e.g.,
Miura maps for the supercurrents of $N=2$ super $W_3$ algebra
\cite{{a190},{a23}}.

There remain some interesting problems with the Boussinesq equation itself.
In particular, it is desirable to have a full understanding of the
relationship with the Hamiltonian formulation and the formulation which uses
the Gel'fand-Dikii brackets. Also it is as yet unknown how to
incorporate in the present scheme in a simple way next equations from the
Boussinesq hierarchy. To this end it seems natural to extend the
coset spaces of $W_3^{\infty}$ by placing
in the coset some higher-spin
generators $J^{(s)}_n$ from the stability subgroup,
e.g. the spin four one
$J_n^{(4)}$, and to introduce additional time variables as the coset
parameters associated with the generators $J^{(s)}_{-s+1}$, e.g.
$J_{-3}^{(4)}$.
New coset fields are expected to be removable by
inverse Higgs
effect, still leaving $u$ and $v$ (or $u_1$ and $v_1$ or $u_0$ and $v_0$) as
the only essential fields of the theory. At the same time, due to the
appearance of extra time variables, the essential fields could obey
higher-order
Boussinesq equations with respect to these variables as a consequence of
appropriate extensions of the covariant reduction procedures employed
above.

Finally, we would like to point out that the covariant
reduction approach invented and applied first in the case
of Liouville theory \cite{a14} mainly for the practical purpose of
constructing higher superextensions of this theory \cite{a15}
now turns out to possess a considerably wider range of applicability. It can be
regarded as a universal tool for treating integrable systems in a
manifestly geometric language of the coset space realizations of appropriate
infinite-dimensional symmetries. The Toda
systems \cite{a12}, Boussinesq and KdV  \cite{a17} hierarchies certainly
admit an adequate geometric description in its framework. It would be of
interest
to consider along similar lines other classical integrable systems,
such as the sine-Gordon and nonlinear Schrodinger equations, and to understand
what are analogs of, say, $W_3^{\infty}$ in all these cases. On the other
hand, one of the problems ahead is to apply our nonlinear realization
techniques to all known $W$ type (super)algebras (e.g. Knizhnik-Bershadsky
superalgebras) and to deduce the integrable
equations associated with them.
So our main goal is to provide a common geometrical basis for
various integrable systems in 1+1 dimensions and the present work should be
regarded as a step in this direction.

\setcounter{equation}{0}


\begin{thebibliography}{99}

\bibitem{a1} A.B. Zamolodchikov, Teor. Mat. Fiz. 65 (1985) 347; \\
     V.A. Fateev and S. Lukyanov, Int. Journ. Mod. Phys. A3 (1988) 507
\bibitem{a2} M. Fukuma, H. Kawai and R. Nakayama,
Int. Journ. Mod. Phys. A6 (1991) 1385
\bibitem{a3} R. Dijgraaf, E. Verlinde and H. Verlinde,
Nucl. Phys. B348 (1991) 435; \\
     E. Verlinde and H. Verlinde,  Nucl. Phys. B348 (1991) 457
\bibitem{a4} A. Bilal and J.L. Gervais, Phys.Lett. B 206 (1988) 412;
     Nucl. Phys. B 314 (1989) 579, 646;\\
     I. Bakas, Nucl. Phys. B 302 (1988) 189
\bibitem{a5} J.L. Gervais, Phys. Lett. B 160 (1985) 277
\bibitem{a6} P. Mathieu, Phys. Lett. B 208 (1988) 101
\bibitem{a7} K. Yamagishi, Phys. Lett. B 259 (1991) 436; \\
	     F. Yu and Y.S. Wu, Phys. Lett. B 236 (1991) 220; \\
     A. Das, W.J. Huang and S. Panda, Phys. Lett. B 271 (1991) 109; \\
     A. Das, E. Sezgin and S.J. Sin, Phys. Lett. B 277 (1992) 435
\bibitem{a8} I.M. Gel'fand and L.A. Dikii, Math. Surv. 30 (5) (1975) 77;
Funkt. Anal. Priloz. 10 (1976) 13; 13 (1979) 13
\bibitem{a9}
V. Drinfel'd and V. Sokolov, J. Sov. Math. 30 (1985) 1975; Sov. Math. Dokl. 23
(1981) 457
\bibitem{a10}
     S. Bellucci and E. Ivanov, Mod. Phys. Lett. A6 (1991) 1269
\bibitem{a11}
     E. Bergshoeff, A. Bilal and K.S. Stelle, Int. J. Mod. Phys. 6 (1991) 5949
\bibitem{a12} E. Ivanov, S. Krivonos and A. Pichugin,
``Nonlinear realizations of $W_3$
symmetry'', preprint JINR E2-91-328, Dubna, 1991
(to appear in Phys. Lett. B)
\bibitem{a13}
S. Coleman, J. Wess and B. Zumino, Phys. Rev. 177 B (1969) 2239; \\
     C. Callan, S. Coleman, J. Wess and B. Zumino,
Phys. Rev. 177 B (1969) 2247; \\
     D.V. Volkov, Sov. J. Part. Nucl. 4  (1973) 3 \\
     V.I. Ogievetsky, in Proceedings of 10th Winter school of
Theoretical Physics
     in Karpach, v.1, p. 117 ( Wroclaw 1974)
\bibitem{a14}
 E.A. Ivanov and S.O. Krivonos, Teor. Mat. Fiz.
58 (1984) 200; Lett. Math. Phys. 8 (1984) 39
\bibitem{a15} E.A. Ivanov and S.O. Krivonos, Lett. Math. Phys. 7 (1983) 523;
     J. Phys. A: Math. Gen. 17 (1984) L671; \\
 E.A. Ivanov, S.O. Krivonos and V.M. Leviant,
Nucl. Phys. B 304 (1988) 601
\bibitem{a16}
 E.A. Ivanov and V.I. Ogievetsky, Teor. Math. Fiz. 25 (1975) 164
\bibitem{a17} E. Ivanov, S. Krivonos and A. Pichugin, in preparation
\bibitem{a18} B. Spence, Phys. Lett. B 276 (1992) 311
\bibitem{a19} E.A. Ivanov, S.O. Krivonos and V.M. Leviant,
J. Phys. A: Math. Gen.
22 (1989) 345
\bibitem{a190} E. Ivanov, S. Krivonos and R.P. Malik, ``$N=2$ super
$W_3^{\infty}$ algebra and $N=2$ Boussinesq equations'',
preprint JINR E2-92-300,
Dubna, July 1992
\bibitem{a20}
H. Lu, C.N. Pope, L.J. Romans, X. Shen and X.-J. Wang,
Phys. Lett. B 264 (1991) 91
\bibitem{a21} K. Schoutens, A. Sevrin and P. van Nieuwenhuizen, Nucl. Phys.
B 349 (1991) 791; \\
C.M. Hull, Nucl. Phys. B 353 (1991) 707
\bibitem{a22}
C.N. Pope, L.J. Romans and X. Shen, Phys. Lett.
B 236 (1990) 173; B 242 (1990) 401;
Nucl. Phys. B 339 (1990) 191; \\
E. Bergshoeff, B. de Wit and M. Vasiliev, Nucl. Phys. B 366 (1991) 315; \\
K.Stelle, ``$w_{\infty}$-Geometry and $w_{\infty}$-Gravity'',
   Preprint JINR E2-91-90, Dubna, 1991; \\
E. Bergshoeff, M.P. Blencowe and K.S. Stelle,
Commun. Math. Phys. 128 (1990) 213;\\
M. Bordemann, J. Hoppe and P. Schaller, Phys. Lett. B 232 (1989) 199;  \\
M.A. Vasiliev, Int. J. Mod. Phys. A6 (1991) 1115
\bibitem{a23} E. Ivanov and S. Krivonos, ``Superfield realizations of
$N=2$ super-$W_3$'', Preprint IC/92/64, Trieste, April 1992

\end{thebibliography}
\end{document}